\documentstyle[sprocl,epsfig]{article}

\def\Journal#1#2#3#4{{#1} {\bf #2}, #3 (#4)}
\def\A{\em ApJ}

\def\AJ{\em AJ}
\def\AA{\em A\&A}
\def\P{\em PASP}
\def\AR{\em ARAA}

\def\be{\begin{equation}}
\def\ee{\end{equation}}
\def\bea{\begin{eqnarray}}
\def\eea{\end{eqnarray}}
\begin{document}
\title{Seyfert Galaxies in Compact Groups}
\author{P. FOCARDI, B. KELM, V. ZITELLI, G. SARTI}
\address{Dipartimento di Astronomia/INAF-OAB, V. Ranzani 1, 41100 Bologna I}
\maketitle\abstracts{ 
We present results concerning the occurrence of Seyfert galaxies
in a new large sample of Compact Groups (Focardi \& Kelm 2002). 
Seyfert galaxies turn out to be relatively rare ($<$ 3\%), with a significant 
dominance of Sy2.  
Seyferts are preferentially associated to Compact Groups displaying 
relatively high velocity dispersion and a large number of neighbours. 
These characteristics, 
together with an excess of ellipticals among companions, suggest that 
Seyferts are to be found preferentially in rich-groups/poor-cluster like CGs. 
} 
\section{Introduction}
Because of their high density (comparable to the galaxy density in clusters) 
and relatively low velocity dispersion ($\approx$ 200-300 $km/s$) 
Compact Groups {\bf (CGs)} are predicted to constitute the most probable 
sites for strong galaxy-galaxy interactions and mergers to occur.
So far, this general expectation has been tested mainly on the Hickson 
Compact Group sample ({\bf HCGs}, Hickson 1982, 1997). 
Indeed, several HCGs show evidence of ongoing interaction, but component 
usually remain distinct, with recognizable morphological type (Sulentic 1997). 
Zepf (1993) has estimated the fraction of currently merging galaxies in 
HCGs to be $\approx$ 7\%, and the fraction of blue ellipticals (which are 
plausible merger remnants) to be similarly low (4 in 55), 
and predominantly associated to faint members (Zepf {\it et al.} 1991).  
Concerning the far infrared (FIR), Hickson {\it et al.} (1989) found the FIR 
emission to be enhanced in HCGs, however Sulentic \& de Mello (1993) and 
Verdes-Montenegro {\it et al.} (1998) 
suggest there is no firm evidence for enhancement.
A similar lack of FIR enhancement is found in the UZC-CG sample 
by Kelm {\it et al.} (2002). These authors also state that data are 
compatible with IRAS galaxies in CGs being plausible candidates for 
accordant redshift projections, rather than interaction triggered starbursting 
galaxies.   

Addressing the issue of AGNs in HCGs, Kelm {\it et al.} (1998) find 
that only $\approx$ 2\% of the HCG member galaxies display a 
Seyfert spectrum.  
They find Sy to be hosted in luminous spirals, as is usually the case. 
However, computing also low-level AGN activity appears to dramatically 
increase the fraction of AGNs in CGs (Coziol {\it et al.} 1998, 
2000), with a significant preference for early type hosts. 
Coziol {\it et al.} (2000) state that AGNs (including low luminosity/dwarf 
sources) are the most frequent (41\%) activity type encountered in CGs. 
A similar high fraction of AGNs in HCGs is retrieved by Shimada {\it et al.} 
(2000), who additionally compare AGN in HCGs with field sources and claim 
that the dense galaxy environment in HCGs does not affect the triggering of 
either AGNs or nuclear starbursts.
\section{Seyfert galaxies in CGs: how common are they?}
We investigate here the occurrence of AGNs in a new large sample of nearby 
Compact Groups (Focardi \& Kelm 2002) identified in a 96\% complete 
flux-limited (2+1)D galaxy catalogue (UZC, Falco {\it et al.} 1999). 
The analysis is restricted to CGs in the radial velocity range 
2500-7500 $km/s$ and to high excitation Sy (type 1 and 2). 
Sy are identified through cross correlation with the Veron-Cetty 
\& Veron (2001) AGN catalogue (V\&V) and/or with the NED database. 
Out of 639 galaxies in 192 CGs only 16 (2.5\%) turn out to be Sy. 
The fraction slightly rises restricting computation 
to spiral hosts (3.3\%) or to the brightest 
(upper quartile) galaxies in CGs (4.4\%). 
The inclusion of Sy3/LINERs would enhance the fraction of AGNs to 4.5\%, 
a value still an order of magnitude lower than estimated by 
Coziol {\it et al.} (2000) or Shimada {\it et al.} (2000). 

Among Sy in CGs only 3 are Sy1. For comparison, in the V\&V catalogue 
the number of Sy1 (within the same redshift range) is only half the 
number of Sy2, thus confirming the paucity of type 1 sources in CGs 
(Coziol {\it et al.} 2000).   
\section{CGs with and without a Seyfert: is there any difference?}
Interactions with companion galaxies are predicted to  
generate instabilities that will possibly lead to the (re)activation 
of an active nucleus. CGs identified in redshift catalogues 
constitute, by definition, an extremely dense environment, however only a 
minor fraction displays a Sy member. 
This means that additional parameters, linked to the exact dynamical status 
of the group and/or to the host galaxy internal parameters do control the  
Sy-triggering mechanism.   
Following, we include data analysis indicating that among CGs those 
resembling rich-group/poor-cluster appear more likely to host a Sy member. 

In {\bf figure 1-left} the velocity distribution of galaxies in 
{\bf CGs with Sy (SyCGs)} and in {\bf CGs without Sy (nonSyCGs)} is displayed. 
The figure shows the distribution of the difference between the observed 
galaxy radial velocity and the mean velocity of galaxies in the group to 
which it belongs.  
The majority of velocities fall below 200 $km/s$ in both samples. 
However it clearly emerges that galaxies in SyCGs are more likely 
(at 3 $\sigma$ c.l. according to the KS test) than those in nonSyCGs to 
display larger velocity differences.    
 
In {\bf figure 1-right} distributions of the number-of-neighbours within 
an region of 1$h^{-1}$$Mpc$ radius and $|$$\Delta$$cz$$|$$<$1000$km/s$ 
from the CG center are shown, tracing the galaxy number density around CGs 
on a scale much larger than the CG scale.      
Statistical analysis indicates that SyCGs are more likely 
(at 3 $\sigma$ c.l.) than nonSyCGs to be associated to  
large number of companion galaxies, i.e. they reside in a denser large-scale 
environment. 
\begin{figure}[h]
\begin{tabular}{cc}
\epsfig{figure=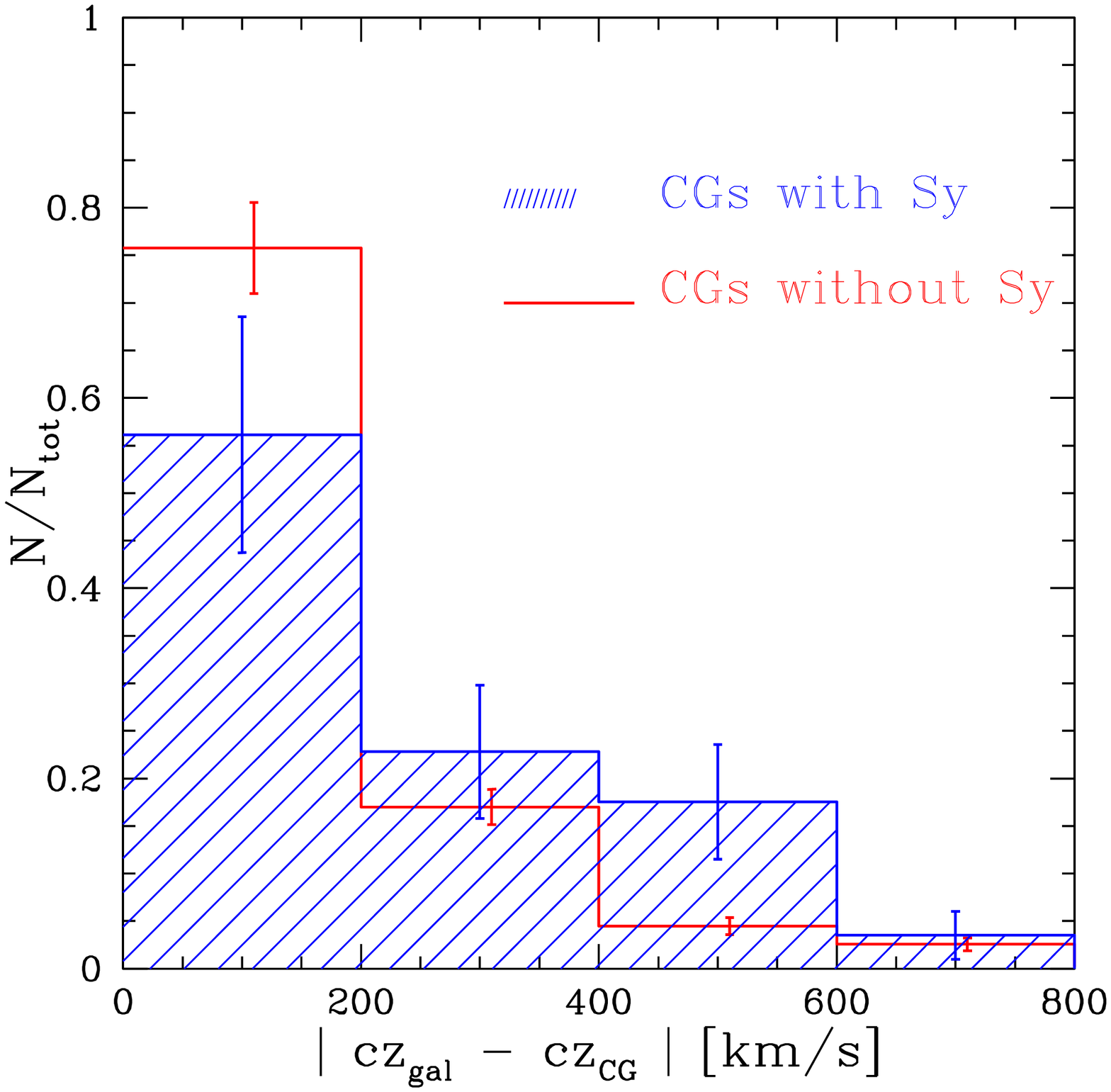, width=5.5cm, height=5.0cm} 
&\epsfig{figure=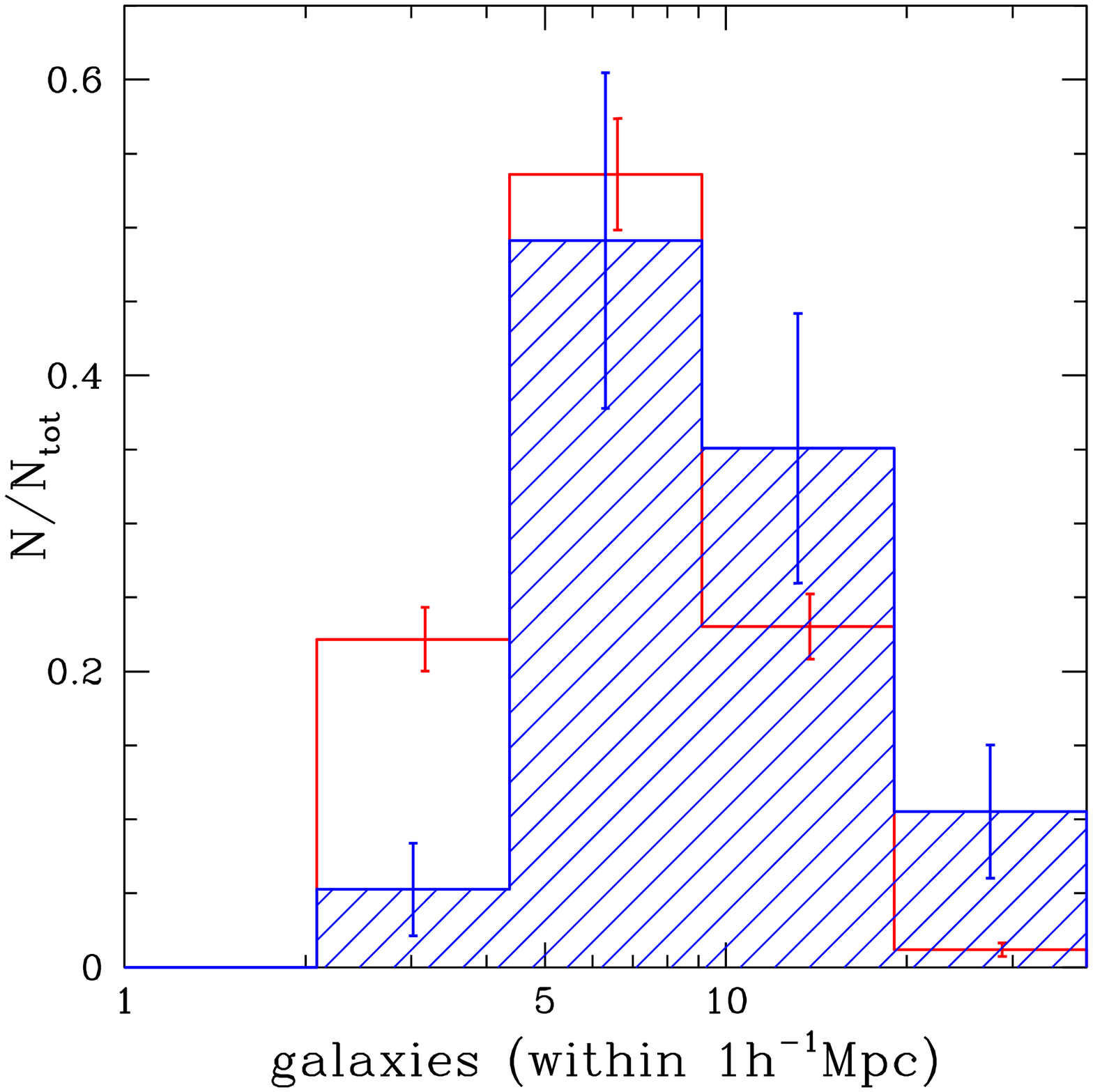, width=5.5cm, height=5.0cm}
\end{tabular}
\caption{Velocity dispersion (left) and number-of-Neighbours (right) 
distributions for CGs hosting/non-hosting a Seyfert. Galaxies in 
SyCGs display higher velocity dispersion and more neighbours than galaxies in nonSyCGs. 
}
\end{figure} 
\begin{figure}
\vskip -0.7 true cm
\centerline{\epsfig{figure=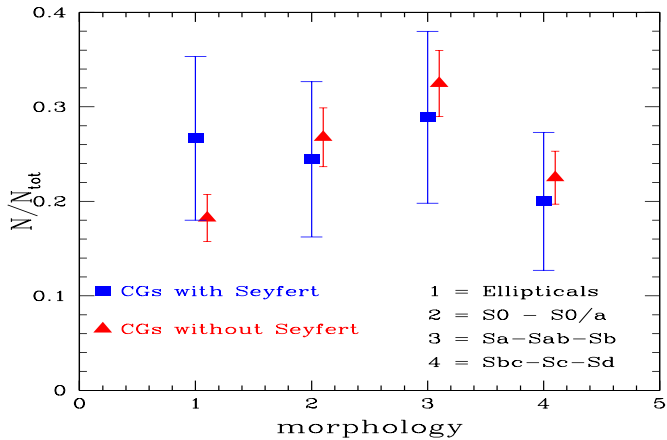, width=9.2cm, height=6.0cm}}
\vskip -0.3 true cm
\caption{Morphology of galaxies in CGs hosting/non-hosting a 
Seyfert member.   }
\end{figure}

In {\bf figure 2} the morphological segregation of galaxies in CGs 
with/without a Sy member are compared. 
Difference points towards 
a population in SyCGs richer in ellipticals. However, among 
Sy themselves only one is an elliptical; accordingly, data indicate that  
Sy are more common when the fraction of ellipticals among companions 
is high. 
\section{Conclusions}
We retrieve only a marginal fraction of Sy galaxies in the UZC-CG 
galaxy sample, suggesting that either 
the interaction-activity connection does not hold for CGs, 
or, that most CGs are actually systems undergoing only mild interactions.
Significant dynamical and environmental differences between CGs 
hosting/non-hosting a Sy member might indicate that Sy are 
typically associated to CGs already resembling rich-groups/poor-clusters. 
This interpretation is also supported by the fact that an excess of 
ellipticals is retrieved among companions to Sy.  
Our analysis clearly suggests that the low fraction of Sy among UZC-CGs 
galaxies results from a low fraction of physical systems in the sample.   
Accordingly, most CGs turn out to be accordant redshift projected groups, or 
groups at first approach in which no strong interaction has yet occurred. 

\end{document}